\documentclass[a4paper,10pt,twocolumn]{article}

\usepackage[left=3cm,right=3cm,top=2.5cm,bottom= 2.5cm]{geometry}
\usepackage{amsmath}
\usepackage{amsfonts}
\usepackage{amssymb}
\usepackage{graphicx}
\usepackage{harvard}
\usepackage{color}
\usepackage{enumerate}
\usepackage{psfrag}
\usepackage{booktabs}
\usepackage[format=hang,margin=10pt,font=small,labelfont=bf]{caption}

\newcommand{\E}[1]{\left\langle #1 \right\rangle}

\definecolor{richtiggelb}{rgb}{0,0.5,0} 
\definecolor{falschrot}{rgb}{0.5,0,0}

\author{M. Weniger, A. Bovier, A. Hense}
\title{A Stochastic Energy Budget Model Using Physically Based Red Noise}

\begin{document}

\twocolumn[
\begin{@twocolumnfalse}

	\maketitle
	
	\begin{abstract}
	A method to describe unresolved processes in meteorological models by physically based stochastic processes (SP) is proposed by the example of an energy budget model (EBM). Contrary to the common approach using additive white noise, a suitable variable within the model is chosen to be represented by a SP. Spectral analysis of ice core time series shows a red noise character of the underlying fluctuations. Fitting Ornstein Uhlenbeck processes to the observed spectrum defines the parameters for the stochastic dynamic model (SDM). Numerical simulations for different sets of ice core data lead to three sets of strongly differing systems. Pathwise, statistical and spectral analysis of these models show the importance of carefully choosing suitable stochastic terms in order to get a physically meaningful SDM.\\
	\end{abstract}
\end{@twocolumnfalse}]

\vspace{1cm}
\section{Introduction}
The atmosphere itself (or the climate system in a wider view) can be seen as a strongly non-linear and infinite-dimensional dynamic system interacting via a multitude of different time- and space-scales of the various flow characterizing variables. For numerical simulations a model system of the atmosphere or the climate system can treat only a finite number of degrees-of-freedom (dof). The effect of the truncated dof on the resolved ones can in general not be neglected due to the scale interactions. Therefore it is necessary to parametrize the effects of the unresolved scales on the resolved ones. In principle the stochastic character of the unresolved scales has to be taken into account. But due to historic reasons most parametrization of subscale processes are deterministic, they model conditional expectation values of moments (mostly second moments) given the resolved scales \cite{Storch97}. A possible problem of numerical weather prediction (NWP) and climate modelling using deterministic parametrization of subscale/unresolved processes is therefore the incomplete consideration of interactions between the resolved and sub-grid scale processes. A stochastic treatment of these parameterizations bears the potential of improving the simulations and providing a better understanding of the stochastic characteristics of the simulated atmospheric or climate system variable. Furthermore a stochastic model yields the natural way to describe the state of a chaotic system, i.e. in form of probability densities. This in turn is closely connected to the analysis of tipping points and regime switches as well as occurrences of extreme events e.g. as revealed in classical statistical physics \cite{Honerkamp02}.\\
The idea of using stochastic climate models was first introduced by Edward \citeasnoun{Lorenz76}: ``\textit{I believe that the ultimate climatic models [...] will be stochastic, i.e., random numbers will appear somewhere in the time derivatives.}'' However in the same article he made a note of caution: ``\textit{If we are truly careful in introducing our random numbers, we can likewise assure ourselves that the probability of producing an ice age, when one ought not to form, is some infinitesimally small number}''. In the same year \citeasnoun{Hasselmann76} published a pathbreaking article which contains the first mathematical formulation of climate models treating weather effects as random forcing terms. Since then there have been various approaches to incorporate stochastic techniques in existing numerical models for atmosphere and climate. In many cases this was proposed or done without the appropriate analyses concerning both the numerical implementation and the mathematical and physical properties of the stochastic variables or processes. While the numerical flaws are easy to avoid in self-contained conceptual models when handled with proper care, it is a non-trivial challenge to implement a stochastic parametrization in an existing more complex model, e.g. a global circulation model (GCM), which consists of deterministic integration schemes. The second aspect is more subtle but equally important: the occurrence and the character of a stochastic parametrization have to be physically justified.\\

The importance of carefully choosing the stochastic terms is the subject of this article. Examplary a energy budget model by \citeasnoun{Budyko69} and \citeasnoun{Sellers69} for the global mean temperature will be considered. Instigated by Hasselmanns paper \cite{Hasselmann76} it was Fraederich, who approached this problem by using stochastic forcing in form of an additive white noise term \cite{Fraedrich78}. Since then it has been an active field of research, see e.g. the review papers \cite{North81,Ghil81}, closely connected to the phenomenon of stochastic resonance \cite{Benzi82}. 

After defining the deterministic framework (\ref{subsec:det-framework}) we will point out problems of the white noise ansatz regarding the model physics (\ref{subsubsec:motivation}). A different, approach using physically based stochastics is proposed with a focus on spectral ice core data analysis (\ref{subsubsec:data-analysis}). It turns out that different data sets lead to three parameter classes for the stochastic terms. For each class one examplary model is considered. The simple mathematical structure of the model exhibits very effective numerical simulations for nonlinear stochastic terms (\ref{subsubsec:numerics}) which in turn allows comprehensive statistical evaluation (\ref{sec:results}). Particular emphasis is placed on the pathwise analysis of the models, since basic statistical properties like mean value and variance are unable to capture the crucial characteristics, i.e. bifurcation structure, timescales of crossings between stable states and correlation between insolation and temperature. After getting an idea of the models behaviour by looking at single realizations (\ref{subsec:samplepaths}), marginal distributions of the temperature at local insolation extrema (\ref{subsec:marginal-distr}) as well as coherence analysis in the frequency domain (\ref{subsec:coherence}) yield solid evidence on the differences between these models. Depending on the choise of ice core data we get fundamentally differing temperature processes. Since there is no reason why a particular data set should be preferred, we come to the conclusion that the model is inapt to be fitted on the type of data available, which is discussed in detail in (\ref{subsec:conclusions}). This however emphasizes the importance of choosing the stochastic terms very careful even in such a simple case. The main focus of this study are not the results of the stochastic EBM but rather the method used to derive them, i.e.
\begin{enumerate}
 \item Identify a variable suited to be describes stochastically
 \item Use spectral analysis of data to determine the distribution of the SP
 \item A rigorous numerical implementation, which will be a nontrivial task in more complex models involving stochastic partial differential equations
 \item Statistical and pathwise analysis of the stochastic dynamic model
\end{enumerate}

\subsection{Terminology}
\label{subsec:terminology}
Looking at various publications regarding stochastic models there is no consistent terminology. We will understand ``Stochastics'' as the field of mathematics which summarizes ``Statistics'' and ``Probability Theory''. Where the later includes, the branches ``Stochastic Processes'' and ``Random Variables''. The term ``Stochastic Parameterization'' will be used to describe models where state independent parameters are replaced by random variables. Models in which time- or state-depending stochastic processes are used will be denoted as ``Stochastic Dynamic Models'' (SDM). On the one hand this emphasizes the fact that the stochastic term itself is part of the systems dynamic and on the other hand it draws the connection to the mathematical field ``Stochastic Dynamics''. Linked to a distinct terminology but with direct mathematical implications is the differentiation of the occuring types of differential equations. Following \cite{Jentzen09b} we will distinguish between the following types of differential equations
\begin{itemize}
 \item deterministic: ordinary differential equations (ODE) / partial differential equations (PDE)
 \item random: RODE / RPDE
 \item stochastic: SODE / SPDE
\end{itemize}
Random differential equations (RODE/RPDE) are ODEs/PDEs with a stochastic process $W_t$ in their vector field, whereas stochastic differential equations (SODE/SPDE) can be characterized by the appereance of differential terms $dW_t$ of a stochastic process, i.e. the unknown quantity itself is a stochastic process. Especially these two classes are often mixed up in meteorological literature. They show essential mathematical differences: RODE/RPDE can be interpreted and analyzed pathwise as their deterministic counterparts. Note that random variables, which can be seen as time constant SPs, are included in the definition of RODE/RPDE. For time dependent SPs however the emerging vector field is usually not differentiable because most driving SPs are only Hoelder continuous in time. This doesn't prohibit the use of the deterministic calculus but leads to lower convergence orders for the numerical schemes \cite{Kloeden07}.\\ 
SODE/SPDE require a different kind of calculus. The vast majority of mathematical literature regarding stochastic differential equations is based on the It\^o calculus \cite{Ito51}, which requires a numerical implementation different from the deterministic one. While techniques for SODEs are well developed, e.g. \cite{Kloeden92}, ``\textit{the numerical solution of SPDEs is at a stage of development roughly similar to that of SODEs in the 1970s}'' \cite{Jentzen09a}. In this article we will have to deal with RODEs and SODEs.

\subsection{A Method for Physically Based Stochastic Models}
\label{subsec:method-SP}
In most cases the starting point for the development of a stochastic model is a deterministic one which captures the gouverning dynamics. The role of stochastic processes is to describe the dynamics of fluctuations and/or uncertainty of model variables and/or sub grid processes with an emphasis on the word ``dynamics'': whereas stochastic parameterizations can be used to represent uncertainties of model parameters, initial values and other time independend variables, SP provide a tool to lay a hand on time and state dependend interactions between subgrid fluctuations and model variables. To refer to the early meaning of stochastic as the ``art of guessing'' it is indispensable to carefully choose the correct SP for a given model. In a more recent article \citeasnoun{Penland08} stated "Simply replacing the fast term with a Gaussian random deviate with standard deviation equal to that of the variable to be approximated, and then using deterministic numerical integration schemes, is a recipe for disaster". While the use of a correct stochastic numerical approach is equally important this paper primarily deals with the problem of finding a physically based SP. As the first step we have to identify unresolved physical processes which may have an effect on the model variables. For this we take a look at each occuring physical variable whose dynamics are not captured by the model, especially the ones which are approximated by a constant value. For two kinds of physical values a stochastic representation is uncalled-for: ``true'' constants, e.g. Stefan-Boltzmann-constant, and non constant variables which could accuratly be described in a deterministic way, e.g. incoming solar radiation depending on Milankovich cycles. The interesting ones are inexact approximations of processes which cannot be described in a deterministic way within the models framework. Of course stochastic representations are also limited by the models space- and time-scales, which provides another criterion in the selection process. When a eligible variable (or a set of eligible variables) is found we need some kind of input on the distribution of the stochastic proccess. This may be information based on models, theoretical insights or data. Deducing the complete distribution of the SP can be a nontrivial task and should be done carefully since even small variations may lead to significant changes in the dynamic structure of the system.

\section{An Energy Budget Model}
\label{sec:EBM}
\subsection{The Deterministic Framework}
\label{subsec:det-framework}
\begin{figure}[hbt]
	\noindent
	\begin{psfrags}%
		\psfragscanon%
		%
		% text strings:
		\psfrag{s03}[t][t]{\color[rgb]{0,0,0}\setlength{\tabcolsep}{0pt}\begin{tabular}{c}T / K\end{tabular}}%
		\psfrag{s04}[b][b]{\color[rgb]{0,0,0}\setlength{\tabcolsep}{0pt}\begin{tabular}{c}Albedo\end{tabular}}%
		%
		% xticklabels:
		\psfrag{x01}[t][t]{270}%
		\psfrag{x02}[t][t]{280}%
		\psfrag{x03}[t][t]{290}%
		\psfrag{x04}[t][t]{300}%
		%
		% yticklabels:
		\psfrag{v01}[r][r]{0.25}%
		\psfrag{v02}[r][r]{0.3}%
		\psfrag{v03}[r][r]{0.35}%
		\psfrag{v04}[r][r]{0.4}%
		%
		% Figure:
		\resizebox{6cm}{!}{\includegraphics{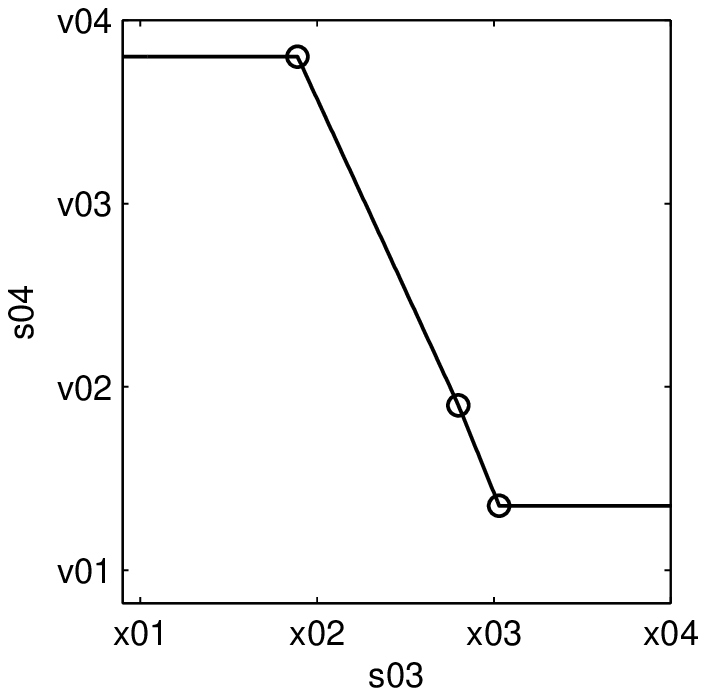}}%
	\end{psfrags}%
	\hspace{2cm}
	\begin{psfrags}%
		\psfragscanon%
		%
		% text strings:
		\psfrag{s03}[t][t]{\color[rgb]{0,0,0}\setlength{\tabcolsep}{0pt}\begin{tabular}{c}Time / ka BP\end{tabular}}%
		\psfrag{s04}[b][b]{\color[rgb]{0,0,0}\setlength{\tabcolsep}{0pt}\begin{tabular}{c}Insolation / $(W/m^2)$ \vspace{5mm}\end{tabular}}%
		%
		% xticklabels:
		\psfrag{x01}[t][t]{0}%
		\psfrag{x02}[t][t]{200}%
		\psfrag{x03}[t][t]{400}%
		\psfrag{x04}[t][t]{600}%
		\psfrag{x05}[t][t]{800}%
		\psfrag{x06}[t][t]{1000}%
		%
		% yticklabels:
		\psfrag{v01}[r][r]{341.5}%
		\psfrag{v02}[r][r]{341.6}%
		\psfrag{v03}[r][r]{341.7}%
		\psfrag{v04}[r][r]{341.8}%
		\psfrag{v05}[r][r]{341.9}%
		\psfrag{v06}[r][r]{342}%
		%
		% Figure:
		\resizebox{6cm}{!}{\includegraphics{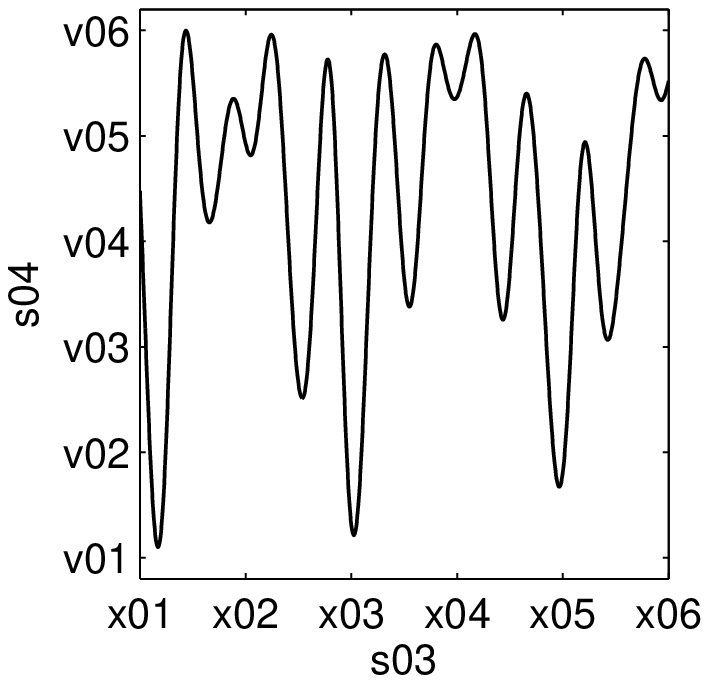}}%
	\end{psfrags}%
	\caption{Albedo parameterization and insolation cycle used in the deterministic framework}
	\label{fig:albedo-insolation}
\end{figure}

The deterministic EBM is given by an ODE for the global mean temperature $T_t$ characterizing the radiation balance between net short wave radiation and outgoing long wave black body radiation
\begin{align*}
	c\ \text{d}T_t &= \big[ S_t(1-\alpha(T)) - \sigma(T_t-\Delta T)^4 \big]\text{d}t.
\end{align*}
Here $c=3*10^8 J/(m^2K)$ denotes the global heat capacity, $S$ the insolation, $\alpha(T)$ the albedo, $\sigma=5.67*10^{-8} W/(m^2K^4)$ the Stefan-Boltzmann-constant and $\Delta T=32.6K$ the temperature offset between surface and the top of the atmosphere. Since we are dealing with a model with spatial dimension zero, obliquity and precession are not accounted for and the solar forcing is solely dependent on the eccentricity of the earth orbit (Fig.\ref{fig:albedo-insolation}). Note that the absolute value of $S$ only fluctuates in the small interval between $341.5K$ and $342.0K$. The albedo was parameterized by deriving the values needed to reproduce stable equilibria at typical ice- and warm-age temperatures, which were estimated from ice core data, and taking into account the present state: $\big(T_1=278.9K, \alpha(T_1)=0.3901\big),\ \big(T_2=288.0K, \alpha(T_2)=0.2950\big),\ \big(T_3=290.3K, \alpha(T_3)=0.2676\big)$. Between these points a stepwise linear interpolation was used (Fig.\ref{fig:albedo-insolation}).\\

Looking at the dynamic structure of this system we observe two stable equilibria at $T_{ice}=278.9K$ and $T_{warm}=290.3K$, which are very unsusceptible to variations in insolation. The temperature value of the third equilibrium strongly depends on the solar forcing and can be found at $T\in (283.5K,287.3K)$. Note that if the insolation would vary in a slightly broader interval, we could observe areas where this equilibrium wouldn't exist, which is why it will become a meta-stable state in the stochastic model. For an arbitrary initial temperature value the system will eventually settle in one of the stable states without the possibilty to cross between these states due to varying insolation.

\subsection{A Stochastic EBM}
\label{subsec:stoch-EBM}
\subsubsection{Motivation}
\label{subsubsec:motivation}
Looking at ice core data immediately reveals, that the temperature is a strongly fluctuating process varying between ice- and warm-ages. The essential missing trait of the deterministic model is the ability to switch between its steady states. One common approach is the use of stochastic forcing in form of an additive white noise term \cite{Fraedrich78}. While being a reasonable first step this ansatz is disputable concerning
\begin{itemize}
 \item \textit{additivity} \\ There is no temperature independent variable in the model which leads to the question what physical process should be described by a temperature independent noise term.
\item \textit{whiteness} \\ Since the physical processes described by this model acts on large space- and timescales it is hard to imagine why a non-correlated noise should adequately capture the occurring fluctuations.
\end{itemize}
This was the motivation to use a different approach. Rather than using a purely mathematical tool to modify the model in order to produce the desired effect, we looked at the variables which are already used by the deterministic EBM to describe one of them as a physically based SP. For a successful stochastic formulation a variable has to satisfy the following conditions: foremost there have to be significant fluctuations of a global mean because the model doesn't resolve spatial dimensions. These fluctuations have to act on a timescale which can be covered by the model, i.e. not smaller than one year. And finally we need evidence on the distribution of the SP. This can be achieved by means of another model resolving the variable in question or ideally through data analysis.

\subsubsection{Data Analysis}
\label{subsubsec:data-analysis}
\begin{figure}[hbt]
	\noindent
	\begin{psfrags}%
		\psfragscanon%
		%
		% text strings:
		\psfrag{s03}[t][t]{\color[rgb]{0,0,0}\setlength{\tabcolsep}{0pt}\begin{tabular}{c}time / ka BP\end{tabular}}%
		\psfrag{s04}[b][b]{\color[rgb]{0,0,0}\setlength{\tabcolsep}{0pt}\begin{tabular}{c}Byrd CO$_2$ / ppm\end{tabular}}%
		%
		% xticklabels:
		\psfrag{x01}[t][t]{20}%
		\psfrag{x02}[t][t]{40}%
		\psfrag{x03}[t][t]{60}%
		\psfrag{x04}[t][t]{80}%
		%
		% yticklabels:
		\psfrag{v01}[r][r]{180}%
		\psfrag{v02}[r][r]{190}%
		\psfrag{v03}[r][r]{200}%
		\psfrag{v04}[r][r]{210}%
		\psfrag{v05}[r][r]{220}%
		\psfrag{v06}[r][r]{230}%
		\psfrag{v07}[r][r]{240}%
		\psfrag{v08}[r][r]{250}%
		\psfrag{v09}[r][r]{260}%
		%
		% Figure:
		\resizebox{6cm}{!}{\includegraphics{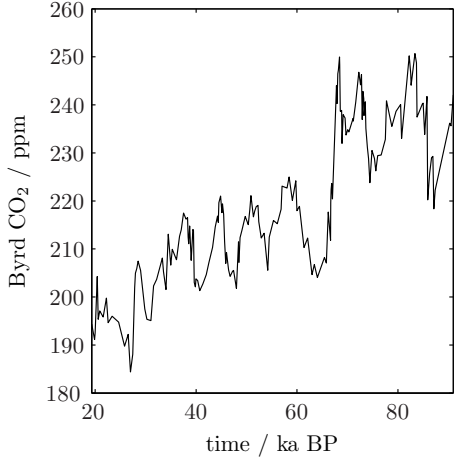}}%
	\end{psfrags}%
	\caption{Byrd ice core data}
	\label{fig:byrd-data}
\end{figure}
Considering these demands we chose $\Delta T$ and its variations due to non constant concentrations of greenhouse gas. We used the following sets of ice core data: Vostok \cite{datavostok}, DomeFuji \cite{datafuji}, Byrd \cite{databyrd}, EPICADome \cite{dataepica}, TaylorDome\cite{datataylor1}, TaylorDome$_2$ \cite{datataylor2}. A graph of the Byrd data (Fig.\ref{fig:byrd-data}) shows examplary that the scaledemands are met and provides direct input regarding the distribution of the SP. One has to be careful on choosing the right data here because there is a strong correlation between CO$_2$ concentration and temperature which involves the danger of predefining the models equilibrium structure. In this simple model we would like to have a SP which describes the ``natural'' small fluctuations of $\Delta T$ excluding feedback effects, i.e. a SP independent of  the model variable $T$, and analyze its effects on the systems dynamic structure. Thus the perfect data would be a large sample of CO$_2$ data on a time interval where the temperature stays constant. Unfortunately this kind of data is not available but we can come close by analyzing detrended dimensionless time-series without crossings between ice- and warm-ages. The first step is the conversion of CO$_2$ data to temperature anomalies, where a doubling in CO$_2$ concentration yields a $\Delta T$ increase of $2K$ and we get
\begin{align*}
	\Delta T' =  \frac{2}{\log{2}}\log{\Big(1+\frac{\text{detrend}(\text{CO}_2)}{\langle\text{CO}_2\rangle}\Big)}.
\end{align*}

\begin{figure}[htb]
	\noindent 
	\begin{psfrags}%
		\psfragscanon%
		%
		% text strings:
		\psfrag{s05}[t][t]{\color[rgb]{0,0,0}\setlength{\tabcolsep}{0pt}\begin{tabular}{c}frequency / pHz\end{tabular}}%
		\psfrag{s06}[b][b]{\color[rgb]{0,0,0}\setlength{\tabcolsep}{0pt}\begin{tabular}{c}intensity / dB\end{tabular}}%
		\psfrag{s10}[][]{\color[rgb]{0,0,0}\setlength{\tabcolsep}{0pt}\begin{tabular}{c} \end{tabular}}%
		\psfrag{s11}[][]{\color[rgb]{0,0,0}\setlength{\tabcolsep}{0pt}\begin{tabular}{c} \end{tabular}}%
		\psfrag{s12}[l][l]{\color[rgb]{0,0,0} fitted OUP}%
		\psfrag{s13}[l][l]{\color[rgb]{0,0,0}Byrd Data}%
		\psfrag{s14}[l][l]{\color[rgb]{0,0,0} 0.95 quantile}%
		\psfrag{s15}[l][l]{\color[rgb]{0,0,0} fitted OUP}%
		%
		% xticklabels:
		\psfrag{x01}[t][t]{0}%
		\psfrag{x02}[t][t]{10}%
		\psfrag{x03}[t][t]{20}%
		\psfrag{x04}[t][t]{30}%
		%
		% yticklabels:
		\psfrag{v01}[r][r]{-50}%
		\psfrag{v02}[r][r]{-45}%
		\psfrag{v03}[r][r]{-40}%
		\psfrag{v04}[r][r]{-35}%
		\psfrag{v05}[r][r]{-30}%
		\psfrag{v06}[r][r]{-25}%
		\psfrag{v07}[r][r]{-20}%
		\psfrag{v08}[r][r]{-15}%
		\psfrag{v09}[r][r]{-10}%
		%
		% Figure:
		\resizebox{6cm}{!}{\includegraphics{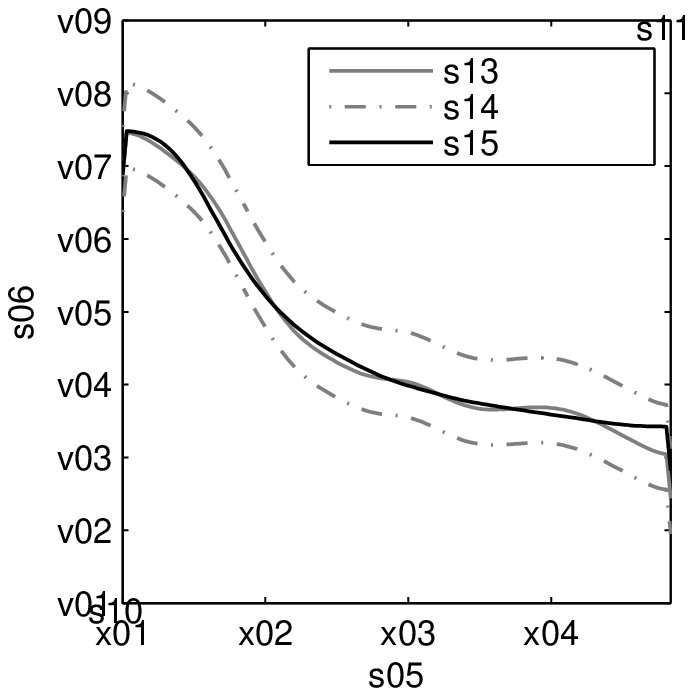}}%
	\end{psfrags}\\%
	%\hspace{2cm}\input{spectrumtaylor.tex}
	\begin{psfrags}%
		\psfragscanon%
		%
		% text strings:
		\psfrag{s05}[t][t]{\color[rgb]{0,0,0}\setlength{\tabcolsep}{0pt}\begin{tabular}{c}frequency / pHz\end{tabular}}%
		\psfrag{s06}[b][b]{\color[rgb]{0,0,0}\setlength{\tabcolsep}{0pt}\begin{tabular}{c}intensity / dB\end{tabular}}%
		\psfrag{s10}[][]{\color[rgb]{0,0,0}\setlength{\tabcolsep}{0pt}\begin{tabular}{c} \end{tabular}}%
		\psfrag{s11}[][]{\color[rgb]{0,0,0}\setlength{\tabcolsep}{0pt}\begin{tabular}{c} \end{tabular}}%
		\psfrag{s12}[l][l]{\color[rgb]{0,0,0} fitted OUP}%
		\psfrag{s13}[l][l]{\color[rgb]{0,0,0}Taylor Data}%
		\psfrag{s14}[l][l]{\color[rgb]{0,0,0} 0.95 quantile}%
		\psfrag{s15}[l][l]{\color[rgb]{0,0,0} fitted OUP}%
		%
		% xticklabels:
		\psfrag{x01}[t][t]{0}%
		\psfrag{x02}[t][t]{20}%
		\psfrag{x03}[t][t]{40}%
		\psfrag{x04}[t][t]{60}%
		\psfrag{x05}[t][t]{80}%
		\psfrag{x06}[t][t]{100}%
		%
		% yticklabels:
		\psfrag{v01}[r][r]{-50}%
		\psfrag{v02}[r][r]{-45}%
		\psfrag{v03}[r][r]{-40}%
		\psfrag{v04}[r][r]{-35}%
		\psfrag{v05}[r][r]{-30}%
		\psfrag{v06}[r][r]{-25}%
		%
		% Figure:
		\resizebox{6cm}{!}{\includegraphics{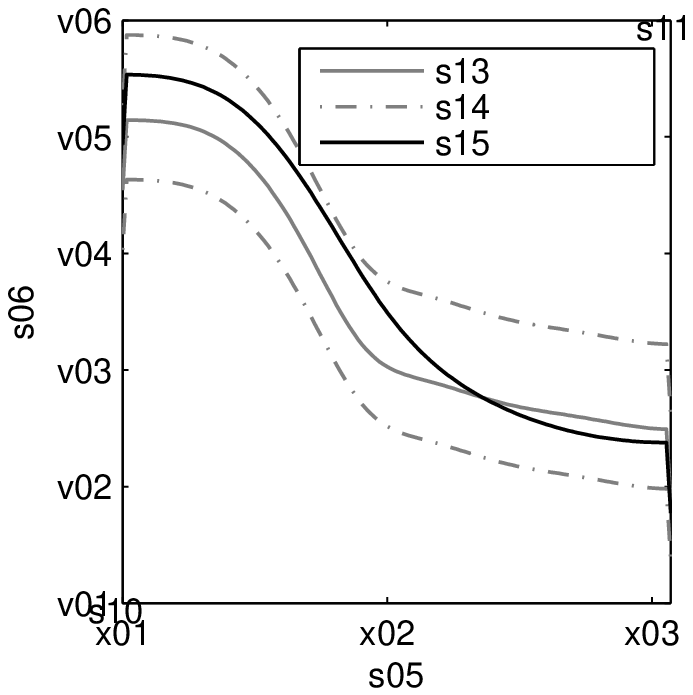}}%
	\end{psfrags}\\%
	%\hspace{2cm}\input{spectrumvostok.tex}\\
	\begin{psfrags}\\%
		\psfragscanon%
		%
		% text strings:
		\psfrag{s05}[t][t]{\color[rgb]{0,0,0}\setlength{\tabcolsep}{0pt}\begin{tabular}{c}frequency / pHz\end{tabular}}%
		\psfrag{s06}[b][b]{\color[rgb]{0,0,0}\setlength{\tabcolsep}{0pt}\begin{tabular}{c}intensity / dB\end{tabular}}%
		\psfrag{s10}[][]{\color[rgb]{0,0,0}\setlength{\tabcolsep}{0pt}\begin{tabular}{c} \end{tabular}}%
		\psfrag{s11}[][]{\color[rgb]{0,0,0}\setlength{\tabcolsep}{0pt}\begin{tabular}{c} \end{tabular}}%
		\psfrag{s12}[l][l]{\color[rgb]{0,0,0} fitted OUP}%
		\psfrag{s13}[l][l]{\color[rgb]{0,0,0}Vostok Data}%
		\psfrag{s14}[l][l]{\color[rgb]{0,0,0} 0.95 quantile}%
		\psfrag{s15}[l][l]{\color[rgb]{0,0,0} fitted OUP}%
		%
		% xticklabels:
		\psfrag{x01}[t][t]{0}%
		\psfrag{x02}[t][t]{2}%
		\psfrag{x03}[t][t]{4}%
		\psfrag{x04}[t][t]{6}%
		\psfrag{x05}[t][t]{8}%
		\psfrag{x06}[t][t]{10}%
		%
		% yticklabels:
		\psfrag{v01}[r][r]{-45}%
		\psfrag{v02}[r][r]{-40}%
		\psfrag{v03}[r][r]{-35}%
		\psfrag{v04}[r][r]{-30}%
		\psfrag{v05}[r][r]{-25}%
		\psfrag{v06}[r][r]{-20}%
		\psfrag{v07}[r][r]{-15}%
		\psfrag{v08}[r][r]{-10}%
		\psfrag{v09}[r][r]{-5}%
		%
		% Figure:
		\resizebox{6cm}{!}{\includegraphics{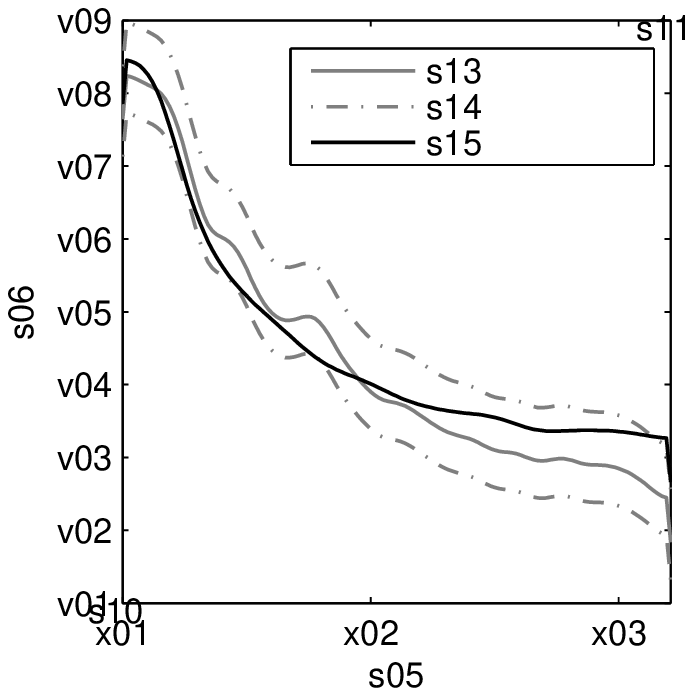}}%
	\end{psfrags}%
	\caption{Spectral analysis of Byrd, Taylor and Vostok ice core data}
\label{fig:ice-core-spectra}
\end{figure}

$\Delta T'$ is the desired data of temperature fluctuations, detrend(CO$_2$) the lineary detrended data from ice core drillings and $\langle\text{CO}_2\rangle$ the mean CO$_2$ level. In order to characterize the SP we did a spectral analyzation of $\Delta T'$ time-series which indicates a red noise distribution (Fig.\ref{fig:ice-core-spectra}). In mathematics this AR(1) process is known as Ornstein-Uhlenbeck process (OUP), or Gauss Markov process, and can be defined by the It\^o SODE
\begin{align*}
	d\Delta T_t &= -\Theta\big( \Delta T_t - \mu \big)\text{d}t + \sqrt{D}\ \text{d}W_t.
\end{align*}
In general the decision between It\^o and Stratonovich calculus is an important one, in this special case however it doesn't make a difference because the noise term is not state dependent. The OUP is defined by three parameters $\mu, \Theta$ and $D$. Since we used dimensionless time-series our approach uses a constant expectation value $\mu$ which is the $\Delta T$ used in the deterministic model. The remaining parameters, which are responsible for the fluctuations, were fitted to the spectral data and confirm the validity of the red noise assumption. Note however that the confidence interval for the parameter values of each time-series is quite large due to low data size. In addition the parameters calculated from different time-series differ strongly and it is not obvious which parameters we should trust:\\

\noindent\begin{tabular}[ht]{lccc} \toprule[1pt]
		\hspace{-2mm}Data & \hspace{-1mm}$\tau [ka]$ & \hspace{-1mm}var $\cdot 10^{-3}$ & \hspace{-1mm}period $[ka]$\\ \midrule[1pt]
		\hspace{-2mm}Vostok & 16.7 & 10.8 & 410\\
		\hspace{-2mm}DomeFuji & 14.0 & 10.6 & 340\\
		\hspace{-2mm}Byrd & \phantom{0}2.2 & 11.2 & \phantom{0}70\\
		\hspace{-2mm}TaylorDome$_2$ & \phantom{0}2.3 & \phantom{0}7.2 & \phantom{0}42\\
		\hspace{-2mm}EPICADome & \phantom{0}2.0 & 11.4 & \phantom{0}12\\
		\hspace{-2mm}TaylorDome & \phantom{0}0.6 & \phantom{0}1.0 & \phantom{0}11\\ \bottomrule[1pt]
\end{tabular}\\

These time-series can be divided into three classes [DomeFuji and Vostok], [Byrd, EPICADome and TaylorDome$_2$] and [TaylorDome]. There are pro and contra arguments for each class: the first one consists of very large data sets covering long time periods. While the high amount of data points is certainly a positive aspect it implies the problem of embedded crossings between ice- and warm-ages. The second class contains the majority of time-series where each member yields similar parameters. The third class consists of only one scarce time-serie with 69 data points. Nevertheless it leads to the weakest noise signal, so one would be ``on the safe side'' using these parameters. Since there is no good argument to exclude either class we analyzed the system for one time-series of each category: Byrd, TaylorDome and Vostok.\\

\subsubsection{Numerical Aspects}
\label{subsubsec:numerics}
The stochastic EBM is a twodimensional system containing one RODE for $T_t$ and one SODE for $\Delta T_t$
\begin{align*}
	c\ \text{d}T_t&= \big[ S_t(1-\alpha(T_t)) - \sigma(T_t-\Delta T_t)^4 \big]\text{d}t \\
	d\Delta T_t &= -\Theta\big( \Delta T_t - \mu \big)\text{d}t + \sqrt{D}\ \text{d}W_t.
\end{align*}
The second equation can be solved analytically yielding
\begin{align*}
	\Delta T_t &= \Delta T_0e^{-\Theta t} + \mu (1-e^{-\Theta t}) \\
		   &\phantom{=\Delta T_0e^{-\Theta t} } + \sqrt{\frac{D}{2\Theta}}e^{-\Theta t}W(e^{2\Theta t}-1).
\end{align*}
This allows the process $\Delta T_t$ to be simulated in a computational cost effective way by generating Gaussian random variables with a variance corresponding to the exponential time scale $e^{2\Theta t}$. The remaining RODE can be solved by common deterministic numerical schemes, allthough they converge at a slower rate, since the process $\Delta T_t$ is not differentiable but only Hoelder continous \cite{Kloeden07}. We used a fourth order Runge-Kutta scheme with a timestep of one year.

\section{Results}
\label{sec:results}
\subsection{Samplepaths}
\label{subsec:samplepaths}

\begin{figure*}[hp!]
	\noindent
	\begin{psfrags}%
		\psfragscanon%
		%
		% text strings:
		\psfrag{s07}[t][t]{\color[rgb]{0,0,0}\setlength{\tabcolsep}{0pt}\begin{tabular}{c}t / ka BP\end{tabular}}%
		\psfrag{s08}[b][b]{\color[rgb]{0,0,0}\setlength{\tabcolsep}{0pt}\begin{tabular}{c}\textbf{Taylor}\\\ T/K\end{tabular}}%
		\psfrag{s09}[t][t]{\color[rgb]{0,0,0}\setlength{\tabcolsep}{0pt}\begin{tabular}{c}t / ka BP\end{tabular}}%
		\psfrag{s10}[b][b]{\color[rgb]{0,0,0}\setlength{\tabcolsep}{0pt}\begin{tabular}{c}\textbf{Byrd}\\\ T/K\end{tabular}}%
		\psfrag{s11}[t][t]{\color[rgb]{0,0,0}\setlength{\tabcolsep}{0pt}\begin{tabular}{c}t / ka BP\end{tabular}}%
		\psfrag{s12}[b][b]{\color[rgb]{0,0,0}\setlength{\tabcolsep}{0pt}\begin{tabular}{c}\textbf{Vostok}\\\ T/K\end{tabular}}%
		%
		% xticklabels:
		\psfrag{x01}[t][t]{0}%
		\psfrag{x02}[t][t]{100}%
		\psfrag{x03}[t][t]{200}%
		\psfrag{x04}[t][t]{300}%
		\psfrag{x05}[t][t]{400}%
		\psfrag{x06}[t][t]{500}%
		\psfrag{x07}[t][t]{600}%
		\psfrag{x08}[t][t]{700}%
		\psfrag{x09}[t][t]{800}%
		\psfrag{x10}[t][t]{900}%
		\psfrag{x11}[t][t]{1000}%
		\psfrag{x12}[t][t]{0}%
		\psfrag{x13}[t][t]{100}%
		\psfrag{x14}[t][t]{200}%
		\psfrag{x15}[t][t]{300}%
		\psfrag{x16}[t][t]{400}%
		\psfrag{x17}[t][t]{500}%
		\psfrag{x18}[t][t]{600}%
		\psfrag{x19}[t][t]{700}%
		\psfrag{x20}[t][t]{800}%
		\psfrag{x21}[t][t]{900}%
		\psfrag{x22}[t][t]{1000}%
		\psfrag{x23}[t][t]{0}%
		\psfrag{x24}[t][t]{100}%
		\psfrag{x25}[t][t]{200}%
		\psfrag{x26}[t][t]{300}%
		\psfrag{x27}[t][t]{400}%
		\psfrag{x28}[t][t]{500}%
		\psfrag{x29}[t][t]{600}%
		\psfrag{x30}[t][t]{700}%
		\psfrag{x31}[t][t]{800}%
		\psfrag{x32}[t][t]{900}%
		\psfrag{x33}[t][t]{1000}%
		%
		% yticklabels:
		\psfrag{v01}[r][r]{279}%
		\psfrag{v02}[r][r]{286}%
		\psfrag{v03}[r][r]{290}%
		\psfrag{v04}[r][r]{279}%
		\psfrag{v05}[r][r]{286}%
		\psfrag{v06}[r][r]{290}%
		\psfrag{v07}[r][r]{279}%
		\psfrag{v08}[r][r]{286}%
		\psfrag{v09}[r][r]{290}%
		%
		% Figure:
		\resizebox{15cm}{!}{\includegraphics{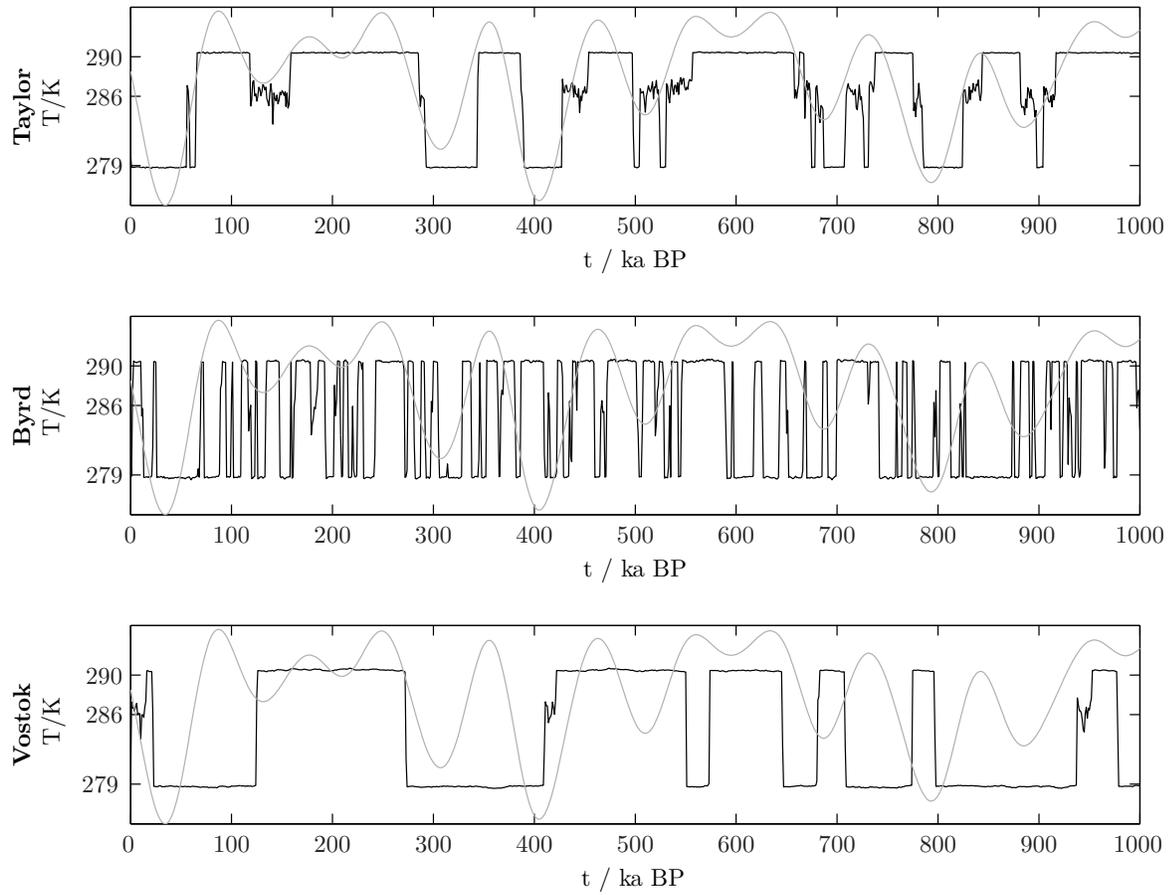}}%
	\end{psfrags}%
	\caption{Samplepaths for temperature (black) and insolation (grey) of three models based on different ice core data.}
	\label{fig:samplepaths}
\end{figure*}
\begin{figure*}[hp!]
	\noindent
	\hspace{-1.5cm}\includegraphics[width=18cm]{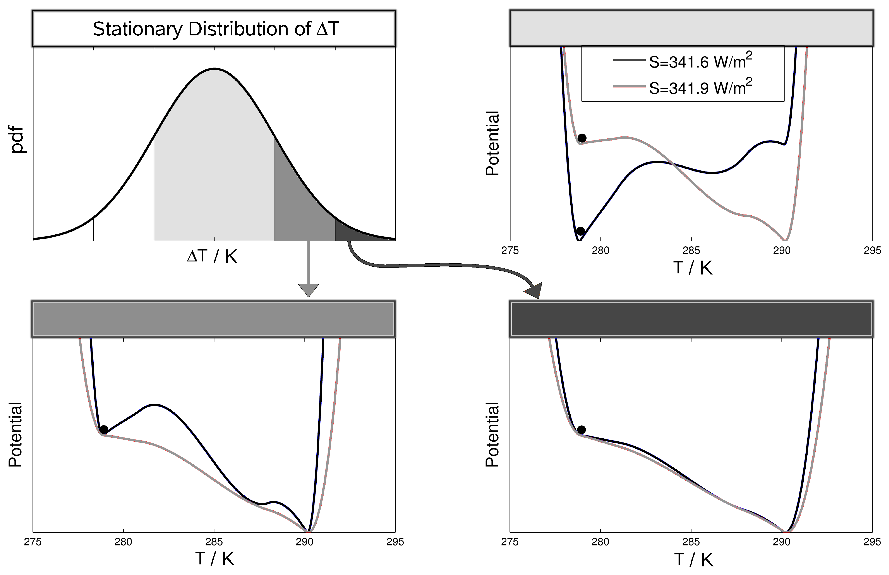}
	\caption{Heuristical explanation of the interaction between solar and stochastic forcing. }
	\label{fig:heuristicpotentials}
\end{figure*}

Looking at a samplepaths (Fig.\ref{fig:samplepaths}) of the three systems gives a first intuitive impression of the models characteristics:
\begin{itemize}
 \item there are at least two stable equilibria coinciding with the deterministic ones
 \item the system can perform jumps between these equilibria
 \item the models differ strongly regarding decorrelationstimes of $T_t$ and strength of correlation between $T_t$ and $S_t$
\end{itemize}

The samplepaths suggests a correlation between solar forcing and temperature depending on the strength of the SP. The coupling of solar and stochastic forcing with the temperature can be explained on a heuristic level, i.e. ignoring the systems inertia, by looking at schematic potential plots (Fig.\ref{fig:heuristicpotentials}). Examplary we will consider positive fluctuations of $\Delta T_t$ where the system (illustrated by a black ball) is in an ice-age. Positive fluctuations of insolation $S_t$ and offset temperature $\Delta T_t$ cause the potential to tilt towards the warm stable state. Since the insolation changes alone are not capable of causing crossings between ice- and warm-ages we have to take a closer look at $\Delta T$. Qualitativly the values of $\Delta T$ can be broken down into three categories:
\begin{itemize}
 \item an area around the mean value
 \item an one-sided area further away from the mean, i.e. a ``weak'' extremum
 \item an one-sided area far away from the mean, i.e. a ``strong'' extremum
\end{itemize}

While $\Delta T$ occupies the first area (light gray) the system stays in its current state unsusceptible to the state of solar forcing, i.e. behaves just like the deterministic system. When $\Delta T$ adopts a weak extremum (medium gray) the system follows the course of insolation, e.g. it jumps from an ice- to a warm-age if and only if the insolation is strong at that point in time. In this case the stochastic forcing amplifies the solar forcing but has no direct effect on the state of the system. This changes  when $\Delta T$ reaches a strong extremum (dark gray): the system will jump regardless of the insolation level, e.g. it will jump from an ice- to a warm-age even if the insolation is minimal. The stochastic forcing is strong enough to cause crossings between stable states even against the effect of the solar forcing. Since the OUP is a Gaussian process, the probabilty for extremes corresponds to $e^{-(\Delta T_t - \langle\Delta T_t\rangle)^2}$. For a system with small fluctuations in $\Delta T_t$, e.g. Taylor with var$(\Delta T_t)\approx10^{-3}K^2$, jumps of the first kind, which are following the insolation, are much more likely than jumps of the second kind. For a system where $\Delta T$ varies strongly, e.g. Byrd or Vostok var$(\Delta T_t)\approx10^{-2}K^2$, crossings of the second kind become more common. Note that the decorrelationtime of $T_t$ corresponds to the solar forcing in the first case but to the stochastic forcing in the second case. This explains the different decorrelationtimes comparing the Byrd and Vostok model.\\

\subsection{Marginal Distributions at Local Extrema of the Solar Forcing}
\label{subsec:marginal-distr}
\begin{figure*}[hbt]
	\noindent
	\begin{psfrags}%
		\psfragscanon%
		%
		% text strings:
		\psfrag{s30}[b][b]{\color[rgb]{0,0,0}\setlength{\tabcolsep}{0pt}\begin{tabular}{c}Strong $S_t$ Min. \\412ka BP\end{tabular}}%
		\psfrag{s31}[t][t]{\color[rgb]{0,0,0}\setlength{\tabcolsep}{0pt}\begin{tabular}{c}T/K\end{tabular}}%
		\psfrag{s32}[b][b]{\color[rgb]{0,0,0}\setlength{\tabcolsep}{0pt}\begin{tabular}{c}\textbf{Taylor}\\ \# realizations\end{tabular}}%
		\psfrag{s33}[b][b]{\color[rgb]{0,0,0}\setlength{\tabcolsep}{0pt}\begin{tabular}{c}Weak $S_t$ Min. \\314ka BP\end{tabular}}%
		\psfrag{s34}[t][t]{\color[rgb]{0,0,0}\setlength{\tabcolsep}{0pt}\begin{tabular}{c}T/K\end{tabular}}%
		\psfrag{s35}[b][b]{\color[rgb]{0,0,0}\setlength{\tabcolsep}{0pt}\begin{tabular}{c}Weak $S_t$ Max. \\849ka BP\end{tabular}}%
		\psfrag{s36}[t][t]{\color[rgb]{0,0,0}\setlength{\tabcolsep}{0pt}\begin{tabular}{c}T/K\end{tabular}}%
		\psfrag{s37}[b][b]{\color[rgb]{0,0,0}\setlength{\tabcolsep}{0pt}\begin{tabular}{c}Strong $S_t$ Max. \\470ka BP\end{tabular}}%
		\psfrag{s38}[t][t]{\color[rgb]{0,0,0}\setlength{\tabcolsep}{0pt}\begin{tabular}{c}T/K\end{tabular}}%
		\psfrag{s39}[t][t]{\color[rgb]{0,0,0}\setlength{\tabcolsep}{0pt}\begin{tabular}{c}T/K\end{tabular}}%
		\psfrag{s40}[b][b]{\color[rgb]{0,0,0}\setlength{\tabcolsep}{0pt}\begin{tabular}{c}\textbf{Byrd}\\ \# realizations\end{tabular}}%
		\psfrag{s41}[t][t]{\color[rgb]{0,0,0}\setlength{\tabcolsep}{0pt}\begin{tabular}{c}T/K\end{tabular}}%
		\psfrag{s42}[t][t]{\color[rgb]{0,0,0}\setlength{\tabcolsep}{0pt}\begin{tabular}{c}T/K\end{tabular}}%
		\psfrag{s43}[t][t]{\color[rgb]{0,0,0}\setlength{\tabcolsep}{0pt}\begin{tabular}{c}T/K\end{tabular}}%
		\psfrag{s44}[t][t]{\color[rgb]{0,0,0}\setlength{\tabcolsep}{0pt}\begin{tabular}{c}T/K\end{tabular}}%
		\psfrag{s45}[b][b]{\color[rgb]{0,0,0}\setlength{\tabcolsep}{0pt}\begin{tabular}{c}\textbf{Vostok}\\ \# realizations\end{tabular}}%
		\psfrag{s46}[t][t]{\color[rgb]{0,0,0}\setlength{\tabcolsep}{0pt}\begin{tabular}{c}T/K\end{tabular}}%
		\psfrag{s47}[t][t]{\color[rgb]{0,0,0}\setlength{\tabcolsep}{0pt}\begin{tabular}{c}T/K\end{tabular}}%
		\psfrag{s48}[t][t]{\color[rgb]{0,0,0}\setlength{\tabcolsep}{0pt}\begin{tabular}{c}T/K\end{tabular}}%
		%
		% xticklabels:
		\psfrag{x01}[t][t]{280}%
		\psfrag{x02}[t][t]{285}%
		\psfrag{x03}[t][t]{290}%
		\psfrag{x04}[t][t]{280}%
		\psfrag{x05}[t][t]{285}%
		\psfrag{x06}[t][t]{290}%
		\psfrag{x07}[t][t]{280}%
		\psfrag{x08}[t][t]{285}%
		\psfrag{x09}[t][t]{290}%
		\psfrag{x10}[t][t]{280}%
		\psfrag{x11}[t][t]{285}%
		\psfrag{x12}[t][t]{290}%
		\psfrag{x13}[t][t]{280}%
		\psfrag{x14}[t][t]{285}%
		\psfrag{x15}[t][t]{290}%
		\psfrag{x16}[t][t]{280}%
		\psfrag{x17}[t][t]{285}%
		\psfrag{x18}[t][t]{290}%
		\psfrag{x19}[t][t]{280}%
		\psfrag{x20}[t][t]{285}%
		\psfrag{x21}[t][t]{290}%
		\psfrag{x22}[t][t]{280}%
		\psfrag{x23}[t][t]{285}%
		\psfrag{x24}[t][t]{290}%
		\psfrag{x25}[t][t]{280}%
		\psfrag{x26}[t][t]{285}%
		\psfrag{x27}[t][t]{290}%
		\psfrag{x28}[t][t]{280}%
		\psfrag{x29}[t][t]{285}%
		\psfrag{x30}[t][t]{290}%
		\psfrag{x31}[t][t]{280}%
		\psfrag{x32}[t][t]{285}%
		\psfrag{x33}[t][t]{290}%
		\psfrag{x34}[t][t]{280}%
		\psfrag{x35}[t][t]{285}%
		\psfrag{x36}[t][t]{290}%
		%
		% yticklabels:
		\psfrag{v01}[r][r]{0}%
		\psfrag{v02}[r][r]{1000}%
		\psfrag{v03}[r][r]{2000}%
		\psfrag{v04}[r][r]{3000}%
		\psfrag{v05}[r][r]{4000}%
		\psfrag{v06}[r][r]{0}%
		\psfrag{v07}[r][r]{1000}%
		\psfrag{v08}[r][r]{2000}%
		\psfrag{v09}[r][r]{3000}%
		\psfrag{v10}[r][r]{4000}%
		\psfrag{v11}[r][r]{0}%
		\psfrag{v12}[r][r]{1000}%
		\psfrag{v13}[r][r]{2000}%
		\psfrag{v14}[r][r]{3000}%
		\psfrag{v15}[r][r]{4000}%
		%
		% Figure:
		\resizebox{15cm}{!}{\includegraphics{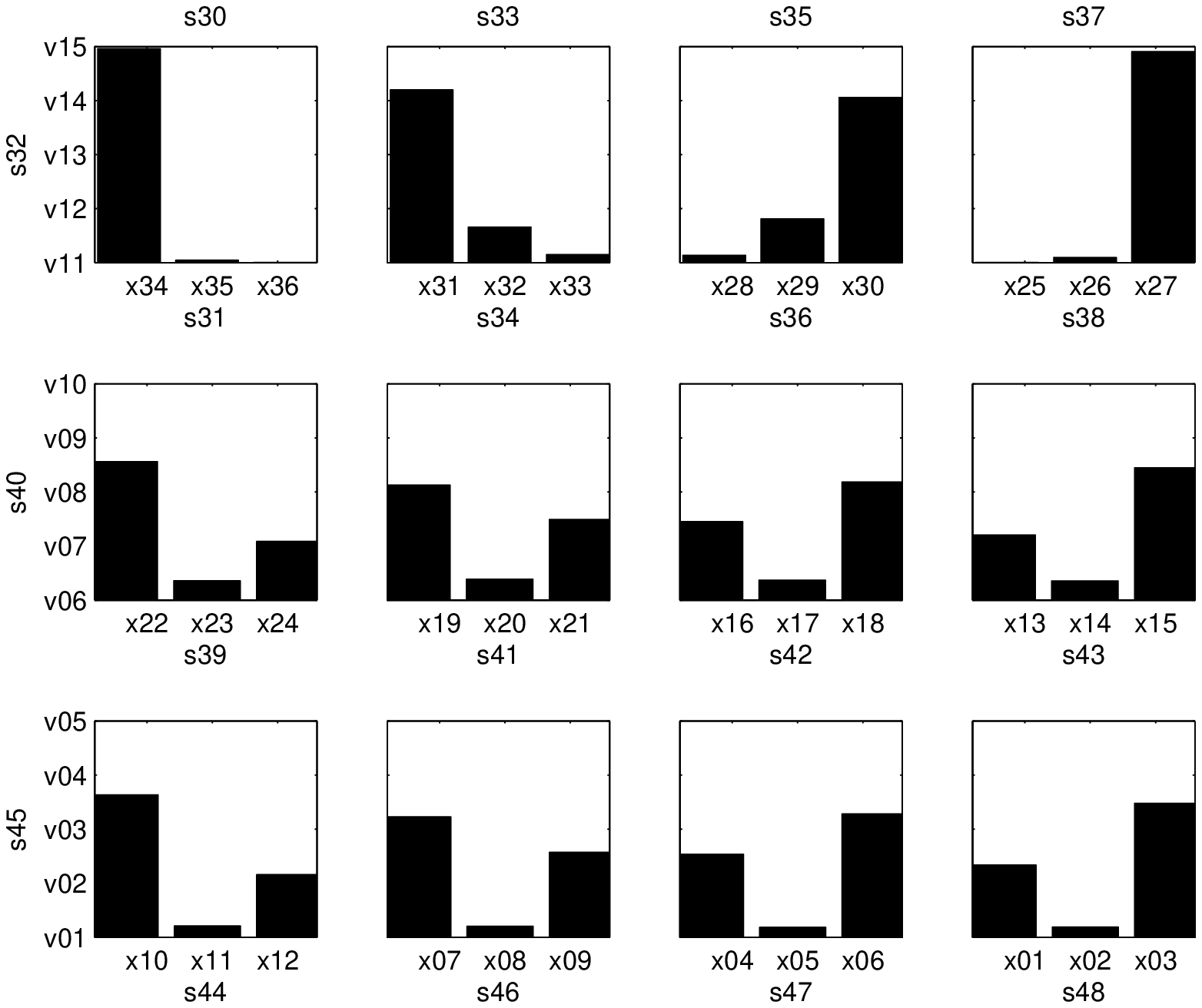}}%
	\end{psfrags}%
	\caption{Marginal distributions at local extrema of the solar forcing $S_t$}
	\label{fig:marginaldistr}
\end{figure*}
Following the intuition gained by samplepaths and heuristic potential considerations we take a first step to describe statistical characteristics of the models by analyzing marginal distributions of $T_t$ at local extrema of the solar forcing $S_t$. We expect the Taylor model to follow the course of insolation and show one-sided distributions, whereas the Byrd and Vostok models should exhibit close to uniform distributions. In Fig.\ref{fig:marginaldistr} the temperature was divided into three intervals corresponding to ice- and warmages and to a state inbetween these two, which is occupied by the system during a crossing. Examplary four local extrema were chosen and the temperature was evaluated with a time lag of $8$ka for 4000 realizations of each model. This value will be derived in the following section \ref{subsec:coherence}. The Taylor model follows our expectations espacially during strong insolation extrema. Even in the case of weaker local extrema the distributions strongly favor the corresponding temperature state. Contrary to the differing samplepaths for the Byrd and Vostok systems, the marginal distributions show a strong resamblance. On the one hand this shows how important a pathwise analysis of a stochastic model is, to capture its characteristics, on the other hand it is an indication that both models have similar correlation between insolation and temperature, independend of the differing decorrelation times. The later show up in form of the differing amount of transition states, which hints at more occuring crossings in the Byrd model.\\

\subsection{Coherence}
\label{subsec:coherence}
\begin{figure*}[htb]
	\noindent 
	\begin{psfrags}%
		\psfragscanon%
		%
		% text strings:
		\psfrag{s07}[t][t]{\color[rgb]{0,0,0}\setlength{\tabcolsep}{0pt}\begin{tabular}{c}Time Lag / ka\end{tabular}}%
		\psfrag{s08}[b][b]{\color[rgb]{0,0,0}\setlength{\tabcolsep}{0pt}\begin{tabular}{c}Crosscorrelation\end{tabular}}%
		\psfrag{s09}[t][t]{\color[rgb]{0,0,0}\setlength{\tabcolsep}{0pt}\begin{tabular}{c}Periodic Time / ka\end{tabular}}%
		\psfrag{s10}[b][b]{\color[rgb]{0,0,0}\setlength{\tabcolsep}{0pt}\begin{tabular}{c}Coherence\end{tabular}}%
		\psfrag{s14}[][]{\color[rgb]{0,0,0}\setlength{\tabcolsep}{0pt}\begin{tabular}{c} \end{tabular}}%
		\psfrag{s15}[][]{\color[rgb]{0,0,0}\setlength{\tabcolsep}{0pt}\begin{tabular}{c} \end{tabular}}%
		\psfrag{s16}[l][l]{\color[rgb]{0,0,0}Vostok}%
		\psfrag{s17}[l][l]{\color[rgb]{0,0,0}Taylor}%
		\psfrag{s18}[l][l]{\color[rgb]{0,0,0}Byrd}%
		\psfrag{s19}[l][l]{\color[rgb]{0,0,0}Vostok}%
		%
		% xticklabels:
		\psfrag{x01}[t][t]{0}%
		\psfrag{x02}[t][t]{30}%
		\psfrag{x03}[t][t]{60}%
		\psfrag{x04}[t][t]{90}%
		\psfrag{x05}[t][t]{120}%
		\psfrag{x06}[t][t]{150}%
		\psfrag{x07}[t][t]{-400}%
		\psfrag{x08}[t][t]{-200}%
		\psfrag{x09}[t][t]{0}%
		\psfrag{x10}[t][t]{200}%
		\psfrag{x11}[t][t]{400}%
		%
		% yticklabels:
		\psfrag{v01}[r][r]{0}%
		\psfrag{v02}[r][r]{0.2}%
		\psfrag{v03}[r][r]{0.4}%
		\psfrag{v04}[r][r]{0.6}%
		\psfrag{v05}[r][r]{0.8}%
		\psfrag{v06}[r][r]{1}%
		\psfrag{v07}[r][r]{-0.4}%
		\psfrag{v08}[r][r]{-0.2}%
		\psfrag{v09}[r][r]{0}%
		\psfrag{v10}[r][r]{0.2}%
		\psfrag{v11}[r][r]{0.4}%
		\psfrag{v12}[r][r]{0.6}%
		\psfrag{v13}[r][r]{0.8}%
		%
		% Figure:
		\resizebox{15cm}{!}{\includegraphics{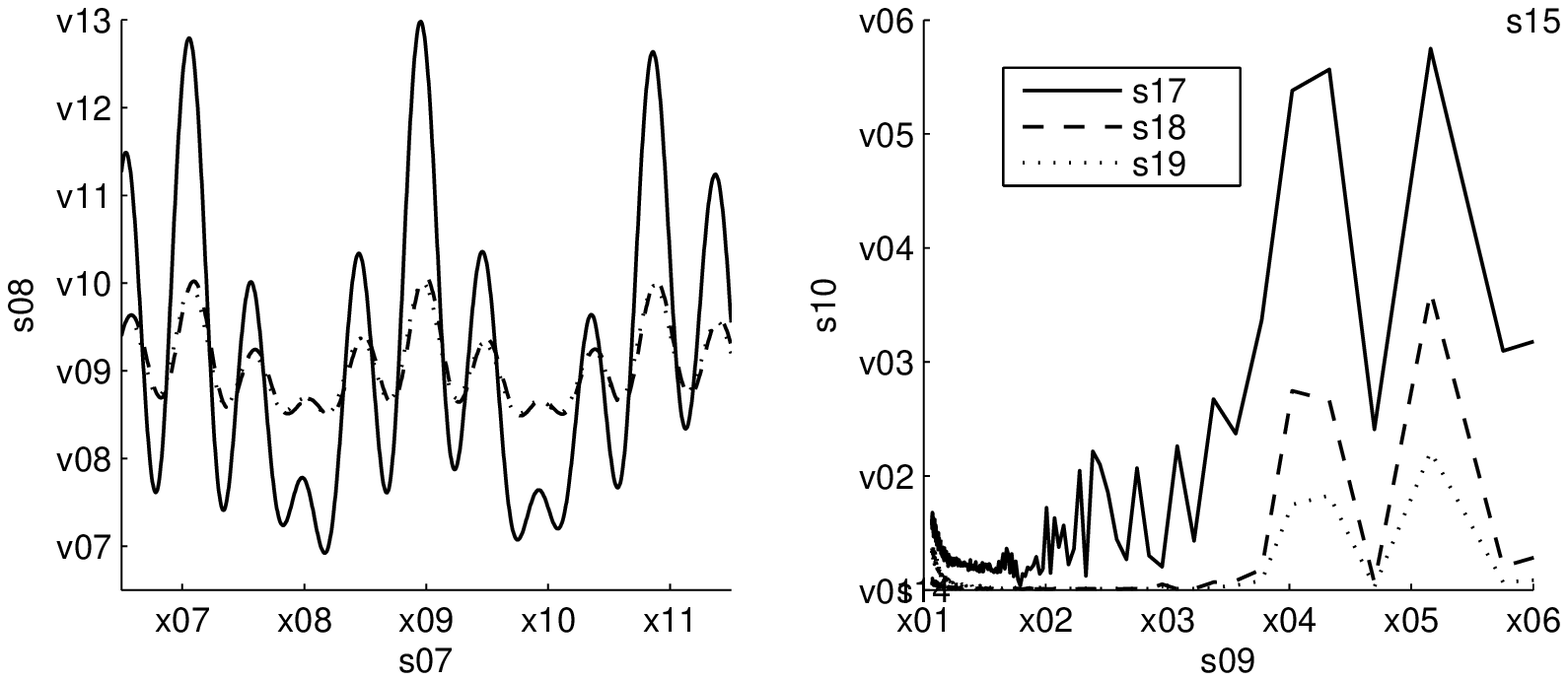}}%
	\end{psfrags}%
\caption{Crosscorrelation and coherence of various ice core data}
\label{fig:xcorr_coherence}
\end{figure*}

Of course neither a single path nor marginal distributions at fixed points in time are sufficient to characterize a stochastic system. Therefore we analyzed the crosscorrelation and coherence between insolation and temperature based on the 4000 realizations of each model we used in the last step. The impression gained thus far are verified by the crosscorrelation (Fig.\ref{fig:xcorr_coherence}), which is much higher in the Taylor system than it is in Byrd in Vostok systems. A maximal crosscorrelation of about $0.8$ at a time lag of $8$ka and the peaks at time lags of about $400$ka, which relates to the eccentric cycle, suggests that the Taylor system is dominated by its insolation. The same periodic structure can be observed for the Byrd and Vostok model but with a much smaller amplitude. In these models a maximal crosscorrelation of about $0.2$ indicates only slight influences of the insolation on the temperature but a dominance of the stochastic forcing.\\
The periodic crosscorrelation structure suggests to take a look at the frequency domain. The (quadratic) spectral coherence of two signals $x,y$ is defined by $C_{xy}^2=|G_{xy}|^2 / G_{xx}G_{yy}$, where $G_{xy}(f)=\E{X^*(f)Y(f)}$ is the cross-spectral density using the Fourier transformed signals $X,Y$. For ergodic linear systems the coherence is a measure for the causality between the input $x$ and the output $y$. Allthough we are not dealing with a linear system, the fact that the insolation $S_t$ is a deterministic signal leads to a closely related interpretation. Denoting the Fourier transforms of insolation and temperature as $\mathcal{S}$ and $\mathcal{T}$ yields
\begin{align*}
	C_{\mathcal{S}\mathcal{T}}^2&=\frac{|\E{\mathcal{S}^*\mathcal{T}}|^2}{\E{|\mathcal{S}|^2}\E{|\mathcal{T}|^2}} =\frac{|\mathcal{S}|^2|\E{\mathcal{T}}|^2}{|\mathcal{S}|^2\E{|\mathcal{T}|^2}}\\
	& =\frac{|\E{\mathcal{T}}|^2}{\E{|\mathcal{T}|^2}} = \Bigg(\frac{\text{var}(\mathcal{T})}{|\E{\mathcal{T}}|^2}+1\Bigg)^{-1}, \hspace{2mm} \E{\mathcal{T}}\neq0.
\end{align*}
The coherence between $S$ and $T$ does not depend on the insolation $S$. It takes on values between $0$ and $1$ and can be interpreted as a measure for the deterministicity of the temperature $T$: a system with no fluctuations has var$(\mathcal{T})=0$ and therefore $C_{T}^2=1$. If the normalized variance is large $C_T$ goes to zero. The frequency interpretation shows the impact of the stochastic forcing on the individual timescales. All models show peaks at periodic times of $100ka$ and $120ka$ (Fig.\ref{fig:xcorr_coherence}), which coincides with the spectrum of the eccentric cycle, i.e. the solar forcing. As expected the peaks for the Taylor model are close to one, whereas Byrd and Vostok models produce significant lower values. 
% ??? Byrd and Vostok: more detailed discussion on coherence

\subsection{Conclusions}
\label{subsec:conclusions}
Although the SP were carefully constructed in each model, the emerging systems differ fundamentely. While the Taylor systems yields the most realistic results, there is no solid ground to utterly dismiss the other data sets. The underlying problem is the discrepance of spatial dimensions between data sets and model space. Ideally we would need data which allows the derivation of global mean fluctuations of $\Delta T$. Another factor are feedback effects between CO$_2$ levels and temperature which are included in the data but not in the model. A third obstacle is the large amount of data points one would need to determine the SP with high accuracy and confidence. When these requirements are met the spectral fitting of SPs onto the data provides a powerful tool, to include unresolved physics into the model. However the diversity of the three systems emphasizes the importance of choosing the stochastic terms very careful even when extending such a simple deterministic model, which puts additive white noise approaches into question.\\

% ???{Outlook}

\end{document}